\documentclass[
    ,final            
  ]
  {aipproc}

\layoutstyle{8x11double}

\begin{document}

\title{Dissipative processes in superfluid neutron stars}


\author{Massimo Mannarelli}{
  address={Departament d'Estructura i Constituents de la Mat\`eria and
Institut de Ci\`encies del Cosmos (ICCUB), Universitat de Barcelona,
Mart\'i i Franqu\`es 1, 08028 Barcelona, Spain}
}

\author{Giuseppe Colucci}{
  address={Universit\`a di
Bari, I-70126 Bari, Italia and .N.F.N., Sezione di Bari,
I-70126 Bari, Italia}
}

\author{Cristina Manuel}{
  address={Instituto de Ciencias del Espacio (IEEC/CSIC),
Campus Universitat Aut\`onoma de Barcelona, Facultat de Ci\`encies, Torre C5
E-08193 Bellaterra (Barcelona), Spain} 
}

\begin{abstract}
We present some results about  a novel  damping mechanism of r-mode oscillations in neutron stars due to processes that change
the number of protons, neutrons and electrons.  Deviations from equilibrium of the  number densities of the various species   lead to the
appearance in the Euler equations of the system of  a dissipative mechanism, the so-called rocket effect.  The evolution of the r-mode oscillations of a  rotating neutron star are influenced by the rocket effect  and we present estimates of the corresponding damping timescales. 
In the description of the system we employ a two-fluid model, with one fluid consisting of all the charged components  locked together by the
electromagnetic interaction,  while the second fluid consists of superfluid neutrons. 
Both components can oscillate however the rocket effect can only efficiently  damp the countermoving  r-mode oscillations, with the two fluids oscillating out of phase. In our analysis we include the mutual friction dissipative process between the neutron superfluid and the charged component. We neglect the interaction between the two r-mode oscillations as well as effects related  with the crust of the star. Moreover, we use a simplified model of neutron star assuming a uniform mass distribution. 
 \end{abstract}

\maketitle


\section{Introduction}
Fermionic superfluidity
might be realized  in the interior of neutron stars where the temperature is low
and the typical energy scale of  particles is extremely high.   In particular,
in the inner crust of standard neutron stars the attractive interaction between neutrons can lead to the formation of a BCS condensate.

A way  to detect or to discard the presence of a superfluid
phase in compact stars
consists in studying the evolution of the r-mode oscillations~\cite{Andersson:2000mf}. R-modes are non-radial
oscillations of a fluid with the Coriolis force acting as the restoring force. They provide a severe limitation on the rotation frequency of a star through coupling to  gravitational radiation. When dissipative phenomena damp these
oscillations the star can rotate without losing  angular momentum to gravitational radiation. If
dissipative phenomena are not strong enough,  the r-mode  oscillations  
will grow exponentially fast in time and the star will keep slowing down until some dissipation mechanism  is able to  damp the r-mode oscillations. In this way one can put some constrains on  the stellar structure ruling out  phases that do not have large enough viscosity.
For such studies it is necessary to consider in detail all the dissipative
processes and the interactions among the various layers of a star.

In Ref. \cite{Colucci:2010wy} we have studied a novel dissipative process associated with the change in the number of protons, neutrons
and electrons. In real neutron stars these processes can take place  in the
outer core and in the inner crust of the star and are related
to beta decays and interactions between the neutron fluid and the crust. 
Both these processes lead to the appearance of a dissipative force (the so-called rocket term) in the Euler equations of the system.
This force is  due to the fact that when two or more fluids  move with different
velocities a  change of one component into the other results in a 
 momentum transfer between the fluids. This change in momentum is not reversible, because it is always the faster moving fluid that will lose momentum.  The resulting dissipative force is  proportional to the mass rate change, and to the relative velocity between the  fluids.
The name ``rocket effect" reminds that the same phenomenon  takes place  
in the dynamical evolution of a rocket whose mass is changing in time as it
consumes its fuel. 

As far as we know, the dissipative force due to the  rocket
term has not been considered in the context of r-mode oscillations. 
In the hydrodynamical equations corresponding to the mass conservation laws, it is in general assumed that the neutron, proton and electron components are separately conserved quantities.
Assuming that the change in the particle densities is due to out of equilibrium direct Urca processes,  we have determined the typical timescale associated with the rocket effect and we have found that it is sufficiently short to damp countermoving r-mode oscillations, with the normal and superfluid component oscillating out of phase.

\section{Rocket effect}

For a system consisting of neutrons,  protons  and  electrons,
the mass conservation law is given by, see {\it
e.g.}~\cite{Prix:2002jn},
\begin{equation}\label{continuity}
 \partial_t\rho_x+\nabla_i(\rho_x v_x^i)= \Gamma_x \ ,
\end{equation}
where $\Gamma_x$ is the particle mass creation rate per unit volume and the 
index $x=n,p,e$ refer to the particle species, that is, neutrons, protons and
electrons. In these equations we have  considered that some process  can
convert neutrons in protons and electrons and {\it vice versa}.
Therefore, we are assuming that the various components are not separately
conserved. 

One possible mechanism leading to a change in 
the particle number densities is the {\it crust-core transfusion},
a process that may take place when  the ionic
constituent of the crust are squeezed by the underlying superfluid and part of their hadronic  content is released and augments the superfluid  component.  The opposite mechanism, related to a reduction of the pressure leads to the  capture of protons and neutrons by the ions of the
crust. 

A different mechanism  leading to a change in 
the particle number densities is  the direct Urca process 
\begin{equation}
n  \rightarrow  p + e^- + {\bar \nu}_e \, , 
\qquad
p + e^- \rightarrow n + \nu_e \,.
\end{equation}
 It was found in Ref.~\cite{Lattimer:1991ib}, that for certain realistic equations of
state the direct Urca processes are  allowed when the star density
exceeds the nuclear saturation density $\rho_0 = 2.8 \times 10^{14} $ g
cm$^{-3}$, and the proton
fraction exceeds the threshold value $x_p^c = \frac 19$.  We shall restrict to
consider such processes which might be realized in the
interior of massive  neutron stars.

The three particle creation rates are not independent quantities,
because charge and baryon number conservations imply that
$\Gamma_e = \Gamma_p=  -\Gamma_n$, meaning that only one creation rate is independent.

It is possible to simplify the treatment of the system considering that 
electrons and protons are locked together by the electromagnetic interaction. Thus 
they move with the same velocity. Moreover, charge neutrality implies
that $n_e = n_p$.
Therefore, electrons and protons can be treated as a single charge-neutral fluid
and henceforth we shall refer to this fluid as the ``charged'' component.
The two fluids have mass densities 
\begin{equation}
\rho_n = m_n\, n_n \qquad {\rm and } \qquad \rho_c = m_n\, n_c \,, 
\end{equation}
where $m_n=m_p + m_e$ and $n_e=n_p=n_c$. The Euler equations obeyed by the two
fluids are given by 
\begin{eqnarray}
\label{Eul-neutron}
 (\partial_t +v_n^j\nabla_j)(v_i^n+\epsilon_n
w_i)+\nabla_i(\tilde{\mu}_n+\Phi)+\epsilon_n w^j\nabla_iv_j^n
& = & 0  \,, \nonumber \\
\label{Eul-proton}
(\partial_t +v_c^j\nabla_j)(v_i^c - \epsilon_c w_i)+\nabla_i(\tilde{\mu}_c+\Phi)
-\epsilon_c w^j\nabla_iv_j^c & = & R_T \nonumber
\end{eqnarray}
where $\Phi$ is the gravitational potential,  $i,j$ label the space components;  we have defined a chemical potential
by mass $\tilde{\mu}_x = \mu_x/m_n$,  and 
 ${\bf w} = {\bf v}_c -{\bf v}_n$ represents the relative velocity between the
two fluids.  The quantities   $\epsilon_n$ and $ \epsilon_c$ are 
the entrainment parameters, that are related to the fact that momenta and
velocities of quasiparticles may not be aligned~\cite{Andreev}. 

The term on the right hand
side of Eq.~(\ref{Eul-proton}) is given by
\begin{equation}
R_T = \left(1
-\epsilon_n -\epsilon_c  \right) \frac{\Gamma_n}{\rho_c} w_i \, 
\end{equation}
and represents the rocket term.  
In the analysis of the possible dissipative mechanisms of star oscillations
this term is usually neglected.  Indeed, it is in general assumed that the
neutron, proton and electron numbers are separately conserved quantities, that
is $\Gamma_p = \Gamma_e =\Gamma_n =0$. In Ref.~\cite{Colucci:2010wy} we have considered the effect of this term and obtained the corresponding damping timescale, assuming that the change in the number densities is due to direct Urca processes.

\subsection{Stability analysis}
In our analysis we have considered two different r-mode oscillations.
One is associated  predominantly with toroidal comoving displacements of the normal and superfluid components of the system. The second type of r-mode oscillation  is dominated by  toroidal countermoving (out of phase) displacements of the two components. We 
refer to these oscillations as ``standard" r-modes and as
``superfluid" r-modes, respectively. These two modes decouple
for a star made by  uniform and incompressible matter, and we shall restrict to treat  such a case. 

As in Ref.~\cite{Haskell:2009fz}, we have studied the linearized hydrodynamical
equations for the perturbations around an equilibrium configuration of a
neutron star rotating with constant angular velocity $\Omega$, 
and we assume that the background configuration is such that the two fluids
move with the same velocity, thus at equilibrium ${\bf w} =0$.

For simplicity we have considered a toy-model neutron star with
uniform density $\rho=2.5 \rho_0$, with radius $R=10$ km and  mass $M\simeq
1.47 M_\odot$, and no crust. 

\begin{figure}[tbp]
\vspace{2cm}
\includegraphics[height=.22\textheight]{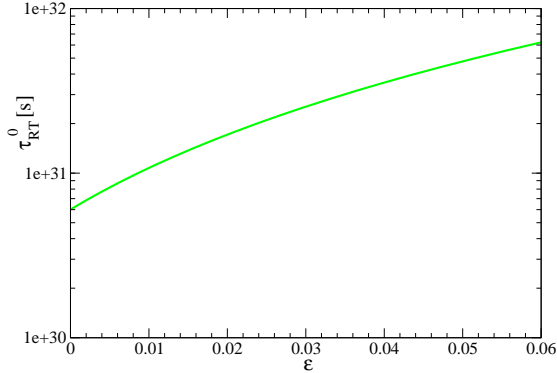}
\caption{ Damping time associated with the rocket term decreases with increasing 
entrainment parameter $\epsilon$, see Ref.~\cite{Colucci:2010wy} for more details.  We have taken $T=10^{9}$K and  a critical  superfluid temperature $T_c=10^{10}$K. }
\label{fig:tau0vsEpsilon}
\end{figure}

In Fig.~\ref{fig:tau0vsEpsilon} we report the damping timescale associated to the rocket effect  as a function of the entrainment parameter $\epsilon$, see Ref.~\cite{Colucci:2010wy}, for standard r-modes. The typical timescale of gravitational wave emission is of the order of tens of seconds, therefore the rocket effect is not able to efficiently damp these oscillations.

\begin{figure}
\includegraphics[height=.22\textheight]{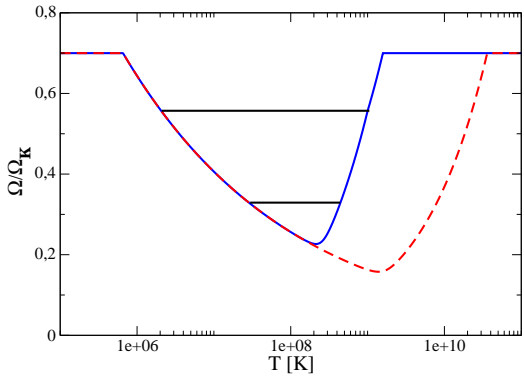}
\caption{Instability window of the superfluid r-modes of a star with
uniform density $\rho=2.5 \rho_0$, with radius $R=10$ km and 
mass $M\simeq 1.47 M_\odot$. The dashed red line represents the
instability window in the absence of the
rocket term. The full blue  line represents the instability window with the
inclusion of the rocket term.
The horizontal full
lines correspond to the effect of the mutual friction for $\epsilon=0.0002$,
lower line, and $\epsilon=0.002$ upper line. In our simplified model of star,
the mutual friction is independent of the temperature. 
}
\label{fig:instability}
\end{figure}
In Fig.~\ref{fig:instability} we report the results for the superfluid r-mode
``instability window" for a star with uniform density, $\rho = 2.5 \rho_0$, and
$R = 10$ km. It is obtained comparing  the various dissipative timescales with the growth timescale of gravitational wave radiation. Stars above the various lines are unstable.
The dashed red line represents the instability
window in the absence of the rocket effect. The region above
the dashed red line is unstable  when
only shear and bulk viscosity damping  are considered.
At low temperature, shear viscosity is the dominant dissipative
mechanism. With increasing temperature shear viscosity is less efficient and
it can damp r-mode oscillations for smaller and smaller values of the
frequency. The behavior of the bulk viscosity is the opposite and starts to damp
r-mode oscillations for temperatures of the order of $10^{10}$ K.
The full blue  line  represents the effect of the rocket term. It has a
behavior qualitatively similar to the bulk viscosity, but it becomes
effective at smaller temperatures. Therefore, the instability window with the
inclusion of the rocket term is much reduced. The full horizontal lines correspond to
the effect of mutual friction, see e.g.~\cite{Lee:2002fp}, for two different values of the entrainment parameter.

\begin{theacknowledgments}
MM thanks the organizer of the conference ``Quark confinement and the hadron spectrum IX" for the kind invitation.
This work has been supported in part by  the INFN-MICINN grant 
with reference number FPA2008-03918E. The work of CM has been supported  by the Spanish  grant FPA2007-60275. The work of MM has been
supported by the Centro Nacional de F\'isica de Part\'iculas, Astropart\`iculas
y Nuclear (CPAN) and by the Ministerio de Educaci\'on y Ciencia (MEC) under
grant  FPA2007-66665 and 2009SGR502.

\end{theacknowledgments}



\bibliographystyle{aipproc}   


\end{document}